\documentclass[aps,pra,twocolumn,floatfix]{revtex4-2}
\usepackage{amsmath,amssymb,graphicx,bbold}
\usepackage{graphicx}  % needed for figures
\usepackage{dcolumn}   % needed for some tables
\usepackage{bm}        % for math
\usepackage{color}
\usepackage{xcolor}

\usepackage{xcolor}
\usepackage{physics}
\usepackage{ulem}
\usepackage{subfigure}
\usepackage{float}
\providecommand{\abs}[1]{\lvert#1\rvert}

\begin{document}

	\title{Radial-angular coupling in self phase modulation with structured light}
	
	\author{M. Gil de Oliveira$^1$, L. J. Pereira$^1$, A. S. Santos$^2$, K. Dechoum$^1$, A. Bramati$^3$, A. Z. Khoury$^1$}
	\email{azkhoury@id.uff.br}
	\affiliation{1- Instituto de F\'{i}sica, Universidade Federal Fluminense, 24210-346 Niter\'{o}i, RJ, Brazil\\
		2- Centro Brasileiro de Pesquisas Físicas, 22290-180, Rio de Janeiro, RJ, Brazil\\
		3- Laboratoire Kastler Brossel, Paris, France}
	\date{\today}
	
	\begin{abstract}
		In this work we study the evolution of an optical vortex undergoing self phase modulation inside 
		a nonlinear Kerr medium. The intensity dependent phase evolution couples the angular and radial 
		degrees of freedom of the input vortex, giving rise to a rich dynamics where new radial modes are 
		created. In the short propagation range, this dynamics is well described by a perturbative approach,  
		predicting the generation of modes with radial numbers between zero and the absolute value of 
		the vortex topological charge. This prediction is confirmed by numerical simulations of the nonlinear 
		propagation equation.
	\end{abstract}
	%\pacs{03.65.Vf, 03.67.Mn, 42.50.Dv}
	%\vskip2pc 
	
	\maketitle
	\section{Introduction}
	\label{sec:intro}
	
	The role played by the orbital angular momentum (OAM) of light in nonlinear optical processes was investigated 
	in seminal works on second harmonic generation \cite{Miles1996-SHG,Miles1997-SHG} and 
	parametric down conversion \cite{Mair2001,Caetano2002}.
	Recently, it has received a renewed attention due to the rich dynamics involved and potential implications in 
	quantum optical processes. As a research field on its own, OAM has become an important tool in optical communications,
	where multiplexing techniques are investigated \cite{Wilner2015,Gong2019,Gregg2019}.
	In centrosymmetric media, such as atomic vapors and glasses, the nonlinear electric susceptibility can only 
	display odd orders \cite{Bloembergen1977}.
	%CITAR LIVRO BOYD
	Self-phase modulation is an important effect associated with the cubic nonlinearity, causing amplitude-phase coupling along the 
	propagation of a sufficiently intense beam. This amplitude-phase coupling affects the beam shape evolution as a result of 
	self-focusing, a remarkable effect also present in the dynamics of Bose-Einstein condensates described by the 
	Gross-Pitaevsky equation in the mean field approximation \cite{BEC-RMP-1999,Mendonca2005}.
	The competition between self-focusing and diffraction may result in morphological stability along propagation and give rise
	to optical solitons.
	%CITAR SOLITONS
	
	Since the early works about OAM transfer in second harmonic generation, other systems have been studied, such as 
	cold atoms \cite{Tabosa1999,Tabosa2003} and optical parametric oscillators \cite{Martinelli2004,Rodrigues2018}. 
	The propagation of structured light in nonlinear media has become an active field of research thanks to numerous techniques 
	for shaping the wavefront of laser beams \cite{Matsumoto2008}. 
	%CLÁSSICO: CITAR NANOPH, OPTICA; INDIANO, KOTLYAR
	Both classical \cite{Buono2014,Geneaux2016,Rao2020,Ruchon2021,Wu2022,Braian2021-Nanoph} and quantum \cite{Vollmer2014,Kerdoncuff2021} 
	optical processes have been studied under structured light driving.
	%QUÂNTICO CITAR GAO, PRL UFF, GERD LEUCHS
	Recently, it has been shown that OAM transfer in nonlinear wave mixing occurs under selection rules coupling the 
	angular and radial degrees of freedom 
	\cite{Pereira2017,Buono2018,Buono2020,Zhi-Han2019,Zhi-Han2020,Oliveira2021,Sonja2021,Vianna2022}.
	Essentially, these rules establish that radial modes are generated from the mixture of counter-rotating beams in up-conversion \cite{Pereira2017,Buono2020}
	or co-rotating beams in stimulated down-conversion \cite{Oliveira2021}. The generation of radial modes in optically thin nonlinear media is essentially	governed by the mathematical properties of products between LG modes \cite{Buono2020,Kotlyar2022}. However, in four-wave mixing in atomic vapors, besides the relative chirality of the interacting beams, the Gouy phase match also plays an essential role due to the optical thickness of the 
	nonlinear medium \cite{Sonja2021}. 
	
	In this work, we investigate theoretically the appearance of radial modes when a single input Laguerre-Gaussian (LG) beam, 
	with topological charge $l_0$ and zero radial order, undergoes self-phase modulation inside a Kerr medium.
	The nonlinear propagation is first described with a perturbative approach, where 
	two distinct regions can be identified according to the Rayleigh distance $z_R$ of the input beam. In the focal zone ($z\ll z_R$), 
	a simple selection rule for radial mode generation is derived, once the waist of the mode basis is properly chosen. 
	New LG modes are created with radial numbers ranging between zero and the absolute value of the input topological charge 
	$l_0\,$. In the diffraction zone ($z\gtrsim z_R$), the nonlinear evolution becomes affected by the Gouy phase match and a complex radial structure takes place.
	
	This work is organized as follows, in Section \ref{sec:propagation} we recall the basic concepts of self-phase modulation in nonlinear Kerr media. In Sections \ref{sec:spmLG} and \ref{sec:perturbativepropagation} we derive a set of coupled nonlinear equations for the mode amplitudes in the Laguerre-Gaussian basis and a perturbative approach to the solution. In Sections \ref{sec:focal zone} and \ref{sec:diffraction zone}, we analyze the radial mode generation using first order perturbation in the focal and diffraction zones, respectively. Section \ref{sec:numerical} is devoted to a numerical analysis, which validates the analytical results presented in the preceding sections. Finally, our conclusions are drawn in Section \ref{sec:conclusion}.	The transformation between Laguerre-Gaussian modes with different waists is presented in Appendix \ref{app:A}. 
	%%%%%%%%%%%%%%%%%%%%%%%%%%%%%%%%%%%%%%%%%%%%%%%%%%%%%%%%%%%%%%%%%%%%%%%%%%%%%%%%%%%%%%%%%%%%%%%%%%%%%%%%%%%%%%%%%%%%%%%%%%%%%%%
	
	\section{Self phase modulation}
	\label{sec:propagation}
	
	Let us consider the propagation of a light beam with frequency $\omega$ undergoing self phase 
	modulation inside a nonlinear Kerr medium with linear index of refraction $n\,$. 
	The wave equation for the complex electric field amplitude $\mathcal{E}(\textbf{r})$ under 
	paraxial propagation is
	\begin{equation}
		\nabla_{\perp}^{2} \mathcal{E}(\textbf{r}) + 
		2ik \frac{\partial\mathcal{E}(\textbf{r})}{\partial z}
		= -\frac{\omega^2}{\varepsilon_{0}c^2} \mathcal{P}^{(3)}(\textbf{r})\,, 
		\label{eq:eqonda}
	\end{equation}
	where $\nabla_{\perp}^{2}$ is the Laplacian operator on the transverse coordinates, 
	$\mathcal{P}^{(3)}(\textbf{r})$ is the nonlinear polarization of the medium and $k = n\omega/c$       is the wave number inside the medium.
	The nonlinear polarization is 
	\begin{equation}
		\mathcal{P}^{(3)}(\textbf{r}) = 3\varepsilon_{0}\chi^{(3)}\abs{\mathcal{E}(\mathbf{r})}^2\mathcal{E}(\mathbf{r})\,,
		\label{eq:P3}
	\end{equation}
	where $\chi^{(3)}$ is the third order nonlinear susceptibility. The nonlinear propagation equation is      obtained by plugging Eq. \eqref{eq:P3} 
	into \eqref{eq:eqonda},
	\begin{equation}
		\nabla_{\perp}^{2} \mathcal{E}(\textbf{r}) + 
		2ik \frac{\partial\mathcal{E}(\textbf{r})}{\partial z} = 
		- 3\chi^{(3)} \frac{\omega^2}{c^2} \abs{\mathcal{E}(\mathbf{r})}^2\mathcal{E}(\mathbf{r}) \,.
		\label{eq:eqondanl}
	\end{equation}
	In the short propagation range, this nonlinear equation gives rise to the well known self-phase modulation effect, in 
	which the field intensity alters the effective index of refraction seen by the input beam. This effect couples the 
	amplitude and phase dynamics inside the nonlinear medium, which affects the spatial properties of an incoming Gaussian
	beam. The transverse variation of the beam intensity produces an effective lens, causing the beam focalization or
	defocalization, according to the sign of the nonlinear susceptibility. In fact, let us assume a solution to Eq. \eqref{eq:eqondanl} of the form
	\begin{equation}
		\mathcal{E}(\mathbf{r}) = \mathcal{A}_0 \,u(\mathbf{r})\, e^{i\theta(\mathbf{r})}\,,    
		\label{eq:spm-initial-solution}
	\end{equation}
	where $u(\mathbf{r})$ is a normalized solution of the paraxial wave equation and $\theta(\mathbf{r})$ is an additional phase 
	term due to the nonlinear response of the medium.
	%CITAR LIVRO
	Under slow transverse variation 
	($\abs{\boldsymbol{\nabla}_{\perp}u}\abs{\boldsymbol{\nabla}_{\perp}\theta}, \abs{u\,\nabla_{\perp}^2\theta}\ll k\,u\,\partial\theta/\partial z$),
	a simple equation can be easily derived for the phase term,
	\begin{eqnarray}
		\frac{\partial\theta}{\partial z} &=& \frac{g}{2k}\,\abs{u(\mathbf{r})}^2 \,,    
		\label{eq:spm-phase-equation}\\
		g &=& 3\chi^{(3)}\mathcal{A}_0^2 \,\frac{\omega^2}{c^2}\,,
		\label{eq:g}
	\end{eqnarray}
	with the following trivial solution 
	\begin{eqnarray}
		\theta(\boldsymbol{\rho},z) &=& \frac{g}{2k}\,\int_0^z \abs{u(\boldsymbol{\rho},z')}^2 dz'\,, 
		\label{eq:spm-phase-solution}
	\end{eqnarray}
	where $\boldsymbol{\rho}$ is the position vector on the plane transverse to the propagation direction $z\,$.
	Therefore, the phase evolution is directly affected by the intensity distribution of the beam, which acts as
	an effective lens and changes the beam focalization, leading to the well known effects of self-focusing and 
	-defocusing. Eq. \eqref{eq:spm-initial-solution} is well-known and has been used in standard descriptions of 
        self-phase modulation. It provides an accurate description of the self-phase modulation phenomenon for 
        interaction lengths much smaller than the Rayleigh distance.
        However, the appearance of radial-angular coupling in the nonlinear propagation of an
        optical vortex cannot be easily seen from this standard approach.
        Moreover, its validity fails when one considers the nonlinear interaction over distances where diffraction 
        becomes important. 
	We next describe a perturbative approach to the nonlinear interaction and investigate the appearance of radial              structures as a result of self-phase modulation in a Laguerre-Gaussian mode. Besides, the perturbative approach 
        can also be used to describe the nonlinear interaction over distances comparable to the Rayleigh length, where diffraction effects become important.
	
	%%%%%%%%%%%%%%%%%%%%%%%%%%%%%%%%%%%%%%%%%%%%%%%%%%%%%%%%%%%%%%%%%%%%%%%%%%%%%%%%%%%%%%%%%%%%%%%%%%%%%%%%%%%%%%%%%%%%%%%%%%%%%%%
	
	\section{Self phase modulation in the Laguerre-Gauss basis}
	\label{sec:spmLG}

	We will describe the field dynamics in the Laguerre-Gaussian (LG) mode basis, where the orbital angular momentum is well defined. 
	In this basis, the electric field amplitude is decomposed as
	\begin{equation}
		\mathcal{E}(\mathbf{r}) = \sum_{pl} A_{pl}(z)u_{pl}(\mathbf{r})\,,    
		\label{eq:decomposicaoLG}
	\end{equation}
	where $A_{pl}(z)$ is a slowly varying mode amplitude. The mode functions are given by
	\begin{equation}
		u_{pl}(\mathbf{r}) = \sqrt{\frac{2}{\pi}}\frac{\mathcal{N}_{pl}}{w(z)}\!
		\left(2\tilde{\rho}_z^2\right)^{\!\!\frac{|l|}{2}} \!
		L_{p}^{|l|}\!\left(2\tilde{\rho}_z^2\right)
		e^{-(1-i\tilde{z})\tilde{\rho}^2}\!e^{il\phi}\!e^{-i\varphi_{pl}}\,,
		\label{eq:modefunctions}
	\end{equation}
	where we adopted the following definitions,
	\begin{eqnarray}
		w(z) &=& w_0 \sqrt{1+\tilde{z}^2}\,,
		\nonumber\\
		\tilde{z} &=& z/z_R\,,
		\quad
		\tilde{\rho}_z = \rho/w(z)\,,
		\nonumber\\
		\varphi_{pl} &=& (2p+|l|+1)\arctan(\tilde{z})\,,
		\label{eq:definitions}\\
		\mathcal{N}_{pl} &=& \sqrt{\frac{p!}{(p+|l|)!}}\,.
		\nonumber
	\end{eqnarray}
	Here $w_0$ is the beam waist, $z_R = \pi w_0^2/\lambda$ is the Rayleigh distance, $p\in\mathbb{N}$ is the radial number
	and $l\in\mathbb{Z}$ is the topological charge of the LG mode. When the expansion given by Eq. \eqref{eq:decomposicaoLG}
	is substituted in the nonlinear wave equation \eqref{eq:eqondanl}, we obtain a set of coupled nonlinear equations 
	for the mode amplitudes,
	\begin{equation}\label{eq:coupled}
		\frac{dA_{pl}}{dz} = i\frac{\omega^2 \chi^{(3)}}{2k c^2}\,
		R_{p\,p_1p_2p_3}^{\,l\,l_1l_2l_3}A_{p_1l_1}A_{p_2l_2}^{*}A_{p_3l_3}\,,
	\end{equation}
	where we adopted the convention of sum over repeated indices. The intermode coupling is mediated by the 
	four-mode overlap integrals over the transverse plane
	\begin{eqnarray}
		R_{p\,p_1p_2p_3}^{\,l\,l_1l_2l_3} (z) \!\!&=&\!\!\! 
		\int u_{pl}^{*}(\textbf{r})u_{p_1l_1}(\textbf{r})u_{p_2l_2}^{*}(\textbf{r})u_{p_3l_3}(\textbf{r})\,d^2\boldsymbol{\rho}\,,
		\nonumber\\
		\label{eq:overlaps}		
	\end{eqnarray}
	which depend on the longitudinal coordinate $z\,$. 
	From these integrals we obtain the selection rules constraining the intermode coupling. When a single LG mode is 
	sent into the nonlinear medium, the overlap integrals determine which modes couple to the input one.
	In this way, the coupled equations of motion describe the onset of new transverse modes according to the 
	selection rules. Interestingly, a finite number of LG modes arise in the short range propagation, which can 
	be described by a perturbative solution of the nonlinear wave equation. However, we will show that this 
	finite structure appears in a modified mode basis with reduced waist parameter.
	%%%%%%%%%%%%%%%%%%%%%%%%%%%%%%%%%%%%%%%%%%%%%%%%%%%%%%%%%%%%%%%%%%%%%%%%%%%%%%%%%%%%%%%%%%%%%%%%%%%%%%%%%%%

        \section{perturbative solution of the nonlinear propagation}
	\label{sec:perturbativepropagation}
	
	Let us assume that a single LG mode with waist $w_0\,$, $p=0$ and $l=l_0$ is sent to the nonlinear medium, 
        so the electric field at the entrance of the medium is 
	\begin{equation}
		\mathcal{E}(\rho,\phi,0) = \mathcal{A}_0 \, u_{0l_0}(\rho,\phi,0) \,.
		\label{eq:E0}
	\end{equation}
	The strength of the nonlinear response is determined by the nonlinear susceptibility and the amplitude of the input field. 
	In this sense, it will be useful to work with the normalized field
	$\psi(\mathbf{r}) = w_0\,\mathcal{E}(\mathbf{r})/\mathcal{A}_0$ and the normalized coordinates 
        $(\tilde{\boldsymbol{\rho}},\tilde{z}) = (\boldsymbol{\rho}/w_0,z/z_R)$  
	to render the nonlinear propagation equation amenable to perturbation
	\begin{equation}
		\tilde{\nabla}_{\perp}^{2} \psi(\mathbf{r}) + 
		4i \frac{\partial\psi(\mathbf{r})}{\partial \tilde{z}} = 
		- g\abs{\psi(\mathbf{r})}^{2}\psi(\mathbf{r}) \,,
		\label{eq:eqondanl-norm}\\
	\end{equation}
	where $g\,$, defined in Eq. \eqref{eq:g}, is the perturbation parameter. 
        We can propose a solution in the form of a power series in $g\,$,
	\begin{equation}
		\psi(\mathbf{r}) = \sum_{n=0}^{\infty} g^n\, \psi_n (\mathbf{r})\,.
		\label{eq:perturbative-solution}
	\end{equation}
	with the initial conditions
	\begin{eqnarray}
		\psi_0(\rho,\phi,0) &=& w_0\,u_{0l_0}(\rho,\phi,0)\,,
		\nonumber\\
		\psi_n(\rho,\phi,0) &=& 0\quad (n>0)\,.
		\label{eq:perturbative-initial-condition}
	\end{eqnarray}
	By using the expansion in Eq.\eqref{eq:eqondanl-norm} and matching the 
	powers of $g$ on both sides, we arrive at a set of inhomogeneous linear equations, the 
	first two being
	\begin{eqnarray}
		\tilde{\nabla}_{\perp}^{2} \psi_0(\mathbf{r}) + 
		4i \frac{\partial\psi_0(\mathbf{r})}{\partial \tilde{z}} &=& 0 \,,
		\label{eq:eqonda-perturbative-0}\\
		\tilde{\nabla}_{\perp}^{2} \psi_1(\mathbf{r}) + 
		4i \frac{\partial\psi_1(\mathbf{r})}{\partial \tilde{z}} &=& 
		- \abs{\psi_0(\mathbf{r})}^{2}\psi_0(\mathbf{r}) \,.
		\label{eq:eqonda-perturbative-1}
	\end{eqnarray}
	The solution of Eq.\eqref{eq:eqonda-perturbative-0} for the zero order term is obviously 
	\begin{equation}
		\psi_0(\tilde{\rho},\phi,\tilde{z}) = w_0\,u_{0l_0}(\tilde{\rho},\phi,\tilde{z})\,.
		\label{eq:perturbative-solution-0}
	\end{equation}
	We can find the first order term by expanding $\psi_1$ in the LG basis,
	\begin{equation}
		\psi_1(\mathbf{r}) = w_0 \sum_{pl} a_{pl}(z) \, u_{pl}(\mathbf{r})\,.
		\label{eq:perturbative-solution-1-LG}
	\end{equation}
	The mode amplitudes satisfy the dynamical equations
	\begin{eqnarray}
		\frac{d a_{pl}}{d \tilde{z}} &=& \frac{i}{4}\, R_{p0}^{ll_0}(z)\,,
		\label{eq:apl-perturbative-1}
		\\
		R_{p0}^{ll_0}(z) &=& w_0^2\int u_{pl}^{*}(\mathbf{r})\,
		U_{0l_0}(\mathbf{r})\,d^2\boldsymbol{\rho}\,,
		\label{eq:overlap}
	\end{eqnarray}
	where $U_{0l_0}(\mathbf{r}) = \left|u_{0l_0}(\mathbf{r})\right|^2 u_{0l_0}(\mathbf{r})\,$,
	and the initial conditions are $a_{pl}(0)=0\,$. 
        Note that the integrand in Eq. \eqref{eq:overlap} depends on the longitudinal 
	coordinate explicitly through the Gouy phase factors, which do not depend on 
        $\boldsymbol{\rho}\,$, and implicitly through the width $w(z)\,$. Since the 
	integral runs over the whole transverse plane, the implicit dependence is washed out and 
	the Gouy phases can be factorized. Therefore, one can easily deduce
        that the overlap integral satisfies
	\begin{equation}
		R_{p0}^{ll_0}(z) = \frac{e^{i\Delta\varphi_{pl}(\tilde{z})}}{1+\tilde{z}^2}\,
		R_{p0}^{ll_0}(0)\,,
		\label{eq:overlap-R(z)-R(0)}
	\end{equation}
	where $\Delta\varphi_{pl}(\tilde{z}) = \varphi_{pl}(\tilde{z}) - \varphi_{0l_0}(\tilde{z})\,$.
        Moreover, due to OAM conservation, $R_{p0}^{ll_0}(0)=0$ for $l\neq l_0\,$,
        so the relevant Gouy phase difference $\Delta\varphi_{pl_0}(\tilde{z}) = 2p\arctan(\tilde{z})$
        depends only on $p\,$.
        Therefore, Eq. \eqref{eq:apl-perturbative-1} can be integrated giving
	\begin{equation}
		a_{pl}(\tilde{z}) = \frac{i}{4}\, \Phi_{p}(\tilde{z})\,R_{p0}^{ll_0}(0)\,,
		\label{eq:apl-solution}
	\end{equation}
        where
	\begin{equation}
		\Phi_{p}(\tilde{z}) = \int_{0}^{\tilde{z}} 
        \frac{e^{i\Delta\varphi_{pl_0}(\tilde{z}')}}{1+\tilde{z}'^{\,2}} \,d\tilde{z}'
		= \frac{1}{2ip}\,\left[1-\left(\frac{1-i\tilde{z}}{1+i\tilde{z}}\right)^{p\,}\right]\,.
	\end{equation}
        is the Gouy phase match function. For $p=0\,$, this expression reduces to 
        $\Phi_0(\tilde{z}) = \arctan(\tilde{z})\,$.
        
	In principle, the perturbative solution involves the contribution of an infinite number of LG modes. 
	However, in the focal zone we can show that the overlap $R_{p0}^{ll_0}(\tilde{z})$ selects a finite
	number of coupled modes, provided the waist parameter of the mode basis is properly adjusted.
	As we show next, the proper waist is reduced with respect to the initial beam waist.
	%%%%%%%%%%%%%%%%%%%%%%%%%%%%%%%%%%%%%%%%%%%%%%%%%%%%%%%%%%%%%%%%%%%%%%%%%%%%%%%%%%%%%%%%%%%%%%%%%%%%%%%%%%%

	\section{Radial mode generation in the focal zone ($\tilde{z} \ll 1$)}
	\label{sec:focal zone}
	
        Up to first order in $\tilde{z}\,$, the solution of Eq. \eqref{eq:eqonda-perturbative-1} reads,
	\begin{equation}
		\psi_1(\mathbf{r}) = \frac{i\tilde{z}}{4}\, U_{0l_0}(\boldsymbol{\rho},0) \,.
		\label{eq:solution-perturbative-1-linear-z}
	\end{equation}
        In the focal zone, where $z\ll z_R$ and $w(z)\approx w_0\,$, the mode functions are 
	approximately independent of the longitudinal coordinate $z\,$, so that 
        $R_{p0}^{ll_0}(z)\approx R_{p0}^{ll_0}(0)\,$. In this case, 
	the cubic term in the overlap integral is reduced to
	\begin{equation}
		U_{0l_0}(\boldsymbol{\rho},0)  = 
		\left(\sqrt{\frac{2}{\pi}}\frac{\mathcal{N}_{0l_0}}{w_0}\right)^{\!\!3}\!\!
		\left(\frac{\sqrt{2} \rho}{w_0}\right)^{\!\!3|l_0|} \!\!
		e^{\!\!-\frac{3\rho^2}{w_0^2}}\,e^{il_0\phi} \,.
		\label{eq:cubicLG}
	\end{equation}
	This term can be written as a superposition of pure LG modes with a rescaled waist $\bar{w}_0$ 
	and Rayleigh distance $\bar{z}_R = \pi\bar{w}_0^2/\lambda\,$.
	This is accomplished by inverting the expression of the Laguerre polynomials in terms of monomials 
    to obtain \cite{Arfken2013}
	\begin{equation}\label{eq:monomial_LG-0}
		\left(\frac{2 \rho^2}{\bar{w}_0^2}\right)^{\!|l_0|} \!\!= 
		|l_0|!\sum_{p = 0}^{|l_0|} \left( \begin{matrix}2|l_0|\\|l_0|-p\end{matrix}\right)
		(-1)^p L_{p}^{|l_0|}\!\left(\frac{2 \rho^2}{\bar{w}_0^2}\right)\,,
	\end{equation}
	where we made $\bar{w}_0 = w_0/\sqrt{3}$ to match the width of the Gaussian term in 
	Eq. \eqref{eq:cubicLG}. In this way, the cubic term can be written as a superposition of LG modes with 
	the rescaled waist $\bar{w}_0\,$,
	\begin{eqnarray}
		&&\!\!\!\!\!\!\!\!U_{0l_0}(\boldsymbol{\rho},0) = \frac{1}{w_0^2}
		\sum_{p = 0}^{|l_0|} C_{pl_0} \,\bar{u}_{pl_0}(\boldsymbol{\rho},0)\,.
		\label{eq:cubicLG-2}\\
		&&\!\!\!\!\!\!\!\!C_{pl_0} = 
		\frac{2\,(-1)^p}{\pi\,(\sqrt{3})^{3\abs{l_0}+1}}
		\left( \begin{matrix}2|l_0|\\|l_0|-p\end{matrix}\right)
		\left( \begin{matrix}|l_0|+p\\p\end{matrix}\right)^{\!\!\frac{1}{2}} ,
		\label{eq:C_{pl}}
	\end{eqnarray}
    where $\{\bar{u}_{pl}\}$ is the set of LG modes with the rescaled waist.
	Therefore, in the focal zone, it will be useful to write the first order 
	solution in terms of these rescaled modes, 
	\begin{eqnarray}
		\psi_1(\mathbf{r}) &=& w_0 \sum_{pl} \bar{a}_{pl}(z) \, \bar{u}_{pl}(\mathbf{r})\,,
		\nonumber\\
		\frac{d \bar{a}_{pl}}{d \tilde{z}} &=& \frac{i}{4}\, \bar{R}_{p0}^{ll_0}(z)\,,  
		\label{eq:perturbative-solution-1-LG-focal}
	\end{eqnarray}
	so the overlap integral becomes
	\begin{equation}
		\bar{R}_{p0}^{ll_0}(z) \approx \bar{R}_{p0}^{ll_0}(0) = \sum_{q = 0}^{|l_0|} C_{ql_0} \!\!
		\int \bar{u}_{pl}^{*}(\mathbf{r}) \,\bar{u}_{ql_0}(\mathbf{r})\,d^2\boldsymbol{\rho}\,.
	\end{equation}
	From the orthonormality of the rescaled LG modes, one trivially obtains
	\begin{eqnarray}
		\bar{a}_{pl} = \frac{i\delta_{ll_0}\tilde{z}}{4}
		\left\{
		\begin{matrix}
			C_{pl_0}\qquad &(p&\leq\,\abs{l_0})\\
			0\qquad &(p& > \,\abs{l_0})
		\end{matrix}
		\right.\,.
	\end{eqnarray}
	Therefore, the final solution up to first order perturbation in the short range is
	\begin{equation}
		\psi(\mathbf{r}) = \psi_0(\mathbf{r}) + \frac{ig\tilde{z}\,w_0}{4}\,
        \sum_{p=0}^{\abs{l_0}} C_{pl_0} \,\bar{u}_{pl_0}(\mathbf{r})\,.
		\label{eq:perturbative-solution-1-LG-focal-final}
	\end{equation}
	Therefore, a finite number of radial modes from $p=0$ to $\abs{l_0}$ arise in the short interaction range.
	The corresponding mode amplitudes grow linearly until diffraction takes place. Note that the same result 
        could be obtained by expanding the standard solution given in Eq. \eqref{eq:spm-initial-solution} up to 
        first order in $g$ and taking the limit $z\ll z_R\,$. However, in the diffraction 
	zone ($\tilde{z}\gtrsim 1$) the Gouy phase match becomes relevant and the self modulated field 
	undergoes a more involved dynamics. In this case, a simple expansion of the standard solution is no 
        longer applicable, while the perturbative approach, including the Gouy phase matching term, provides the 
        proper description.
	%%%%%%%%%%%%%%%%%%%%%%%%%%%%%%%%%%%%%%%%%%%%%%%%%%%%%%%%%%%%%%%%%%%%%%%%%%%%%%%%%%%%%%%%%%%%%%%%%%%%%%%%%%%

	\section{Nonlinear propagation in the diffraction zone ($z \gtrsim z_R$)}
	\label{sec:diffraction zone}
	
	In the diffraction zone, the prefactor appearing in Eq. \eqref{eq:overlap-R(z)-R(0)} cannot 
        be neglected. Moreover, the transverse overlap at $z=0$ is
	\begin{eqnarray}
		R_{p0}^{ll_0}(0) \!\!&=&\!\! w_0^2\int u_{pl}^{*}(\boldsymbol{\rho},0)\,
		U_{0l_0}(\boldsymbol{\rho},0)\,d^2\boldsymbol{\rho}
        \nonumber\\
		\!\!&=&\!\! \sum_{q = 0}^{|l_0|} C_{ql_0} \,
        \int u_{pl}^{*}(\boldsymbol{\rho},0)\,\bar{u}_{ql_0}(\boldsymbol{\rho},0)
        \,d^2\boldsymbol{\rho}\,.
        \label{eq:overlap-diff}
	\end{eqnarray}
	Note that the integral appearing in Eq. \eqref{eq:overlap-diff} 
	is simply the coefficient $\Lambda^{l}_{qp}(3)$ of the mode waist transformation 
        between the bases $\{u_{qm}\}$ and $\{\bar{u}_{qm}\}$
	derived in Appendix \ref{app:A}. 
        Using this result in Eq. \eqref{eq:overlap-diff}, we find
	\begin{eqnarray}
		R_{p0}^{ll_0}(0) &=& 
		\delta_{ll_0} \sum_{q=0}^{\abs{l_0}} \,\Lambda^{l_0}_{qp}(3)\,C_{ql_0}\,.
		\label{eq:R(z)}
	\end{eqnarray}
        This allows us to find the solution of the dynamical equations \eqref{eq:apl-perturbative-1} 
	for the expansion coefficients in the diffraction zone, 
	\begin{equation}\label{eq:apl-perturbative-diffraction-zone}
		a_{pl}(\tilde{z}) = \frac{i\delta_{ll_0}}{4}\,\Phi_{p}(\tilde{z})
        \sum_{q = 0}^{|l_0|} C_{ql_0} \,\Lambda_{qp}^{l_0}(3)\;,
	\end{equation}
	so the first order perturbaton term becomes
	\begin{equation}\label{eq:perturbativesolutionE2}
		\psi(\mathbf{r}) = \psi_0(\mathbf{r}) \!+\! \frac{ig\, w_0}{4}\sum_{q = 0}^{|l_0|} \!C_{ql_0}\!
		\sum_{p=0}^{\infty} \Phi_{p}(\tilde{z})\,\Lambda_{qp}^{l_0}(3)\,u_{pl_0}(\textbf{r}) \,.
	\end{equation}
 	The short range solution Eq. \eqref{eq:perturbative-solution-1-LG-focal-final} is 
        immediately recovered by making $\tilde{z}\ll 1\,$, so that 
	$\Phi_{p}(\tilde{z})\approx \tilde{z} \,\,\forall\,\, (p,l)\,$, and noting that  
	\begin{equation}
		\bar{u}_{ql_0}(\textbf{r}) =  
		\sum_{p=0}^{\infty} \Lambda_{qp}^{l_0}(3)\,u_{pl_0}(\textbf{r})\,.
	\end{equation}
	We next compare the analytical results for the radial mode generation with numerical solutions of the 
	nonlinear propagation equation.
	%%%%%%%%%%%%%%%%%%%%%%%%%%%%%%%%%%%%%%%%%%%%%%%%%%%%%%%%%%%%%%%%%%%%%%%%%%%%%%%%%%%%%%%%%%%%%%%%%%%%%%%%%%%%%%%%%
	
	\section{Numerical Analysis}
	\label{sec:numerical}
	
	Let us now turn to the numerical solution of the nonlinear wave equation \eqref{eq:eqondanl-norm} 
        and compare the behavior predicted for the short range propagation with the perturbative solutions 
        described in the preceding sections.

    In order estimate $g$ in physically realisable cases, we note that
    \begin{equation}
        3\chi^{(3)} \left| \mathcal{E}(\textbf{r}) \right|^2 = 2n_0n_2 I(\mathbf{r}),
    \end{equation}
    where $I(\mathbf{r})$ is the intensity of the beam and $n_0$ ($n_2$) is the linear (nonlinear) refractive index, such that the total refractive index is given by $n = n_0 + n_2I$ \cite{boyd2020nonlinear}. 
    Integrating both sides of this equation over a transverse plane, we get
    \begin{equation}
        3\chi^{(3)} \mathcal{A}_0^2 = 2n_0n_2 P,
    \end{equation}
    where $P$ is the power of the beam. Therefore, we have
    \begin{equation}
        g = 3\chi^{(3)} \frac{\omega^2}{c^2} \mathcal{A}_0^2 = \frac{8\pi^2n_0n_2}{\lambda^2}P\,,
    \end{equation}
    where $\lambda$ is the vacuum wavelength. We can define
    \begin{equation}
        P_{NL} = \frac{\lambda^2}{8\pi^2n_0n_2}\,,
    \end{equation}
    which works as a characteristic power for the nonlinear process. In table \eqref{tab:characteristic power} we show 
    three examples of $P_{NL}$ with different orders of magnitude.
    It is then safe to assume that $g$ can span values from $10^{-6}$ to $10\,$.
    \begin{table}[h]
        \centering
        \begin{tabular}{|c|c|}
        \hline
        \textbf{Medium} & $P_{NL}$ (W) \\
        \hline
        \hline
         Fused silica & $65$ \\
         \hline
         GaAs crystal & $2.7 \times 10^{-2}$ \\
         \hline
         Gold nanoparticles in glass & $7.8 \times 10^{-5}$ \\
         \hline
    \end{tabular}
        \caption{Characteristic power for some systems. Calculated from table 4.1.2 of \cite{boyd2020nonlinear} with $\lambda = 1550 \ nm$.}
        \label{tab:characteristic power}
    \end{table}

    The numerical integration was done using a type 2A step-splitting algorithm \cite{bao2003numerical,chin2020structure}, which is implemented by the package StructuredLight.jl \cite{Oliveira_StructuredLight_jl}. The code used to produce the results of this paper can be accessed at \cite{Oliveira_RadialAngularCouplingKerr}.
    
    First, we get a numerical solution of \eqref{eq:eqondanl-norm}, then we compute the quantity
    \begin{equation}
        \delta \psi = \frac{\psi - \psi_0}{g} = \psi_1 + \order{g}\,,
        \label{eq:deltapsi}
    \end{equation}
    and the projections
    \begin{equation}
    \label{eq:numerical overlap}
        c_{pl_0} = \frac{1}{w_0} \int  \overline{u}_{pl_0}^{\,*}\left(\tilde{\mathbf{r}}\right)\,
        \delta\psi\left(\tilde{\mathbf{r}}\right)  \, d^2 \boldsymbol{\rho}\,.
    \end{equation}
    According to Eq. \eqref{eq:perturbativesolutionE2}, up to first order perturbation, we should have
    \begin{equation}
    \label{eq: cpl nonlinear}
        c_{p l_0} \approx 
        \frac{i}{4} \sum_{q = 0}^{\left | l_0 \right|} C_{ql_0}
        \sum_{r=0}^{\infty} \Phi_{r}(\tilde{z})\,\Lambda_{pr}^{l_0}(3) \Lambda_{qr}^{l_0}(3)\,,
    \end{equation}
    where the summation over $r$ will actually be truncated at some sufficiently large value 
    $r_{max}$ for numerical comparison.
    Note that, for $\tilde{z} \ll 1\,$, we have $\Phi_{r}(\tilde{z})\approx\tilde{z}\,$. Besides, 
    as a basis transformation, $\Lambda_{pr}^{l_0}(\eta)$ must be an orthogonal matrix, so that 
    $\sum_{r} \Lambda_{pr}^{l_0}(\eta) \Lambda_{qr}^{l_0}(\eta) = \delta_{pq}\,$, which leads to
    \begin{equation}
    \label{eq: cpl linear}
        c_{pl_0} \approx \frac{i}{4} \tilde{z}\,C_{pl_0}\,.
    \end{equation}
    \begin{figure}[ht]
        \includegraphics[width=\linewidth]{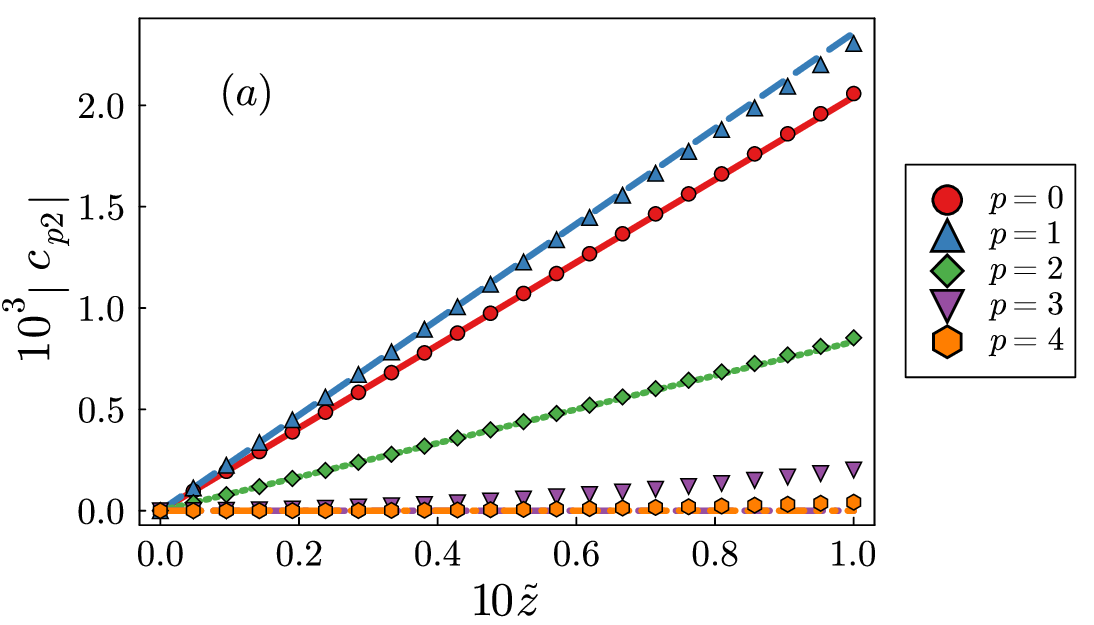}
        \includegraphics[width=\linewidth]{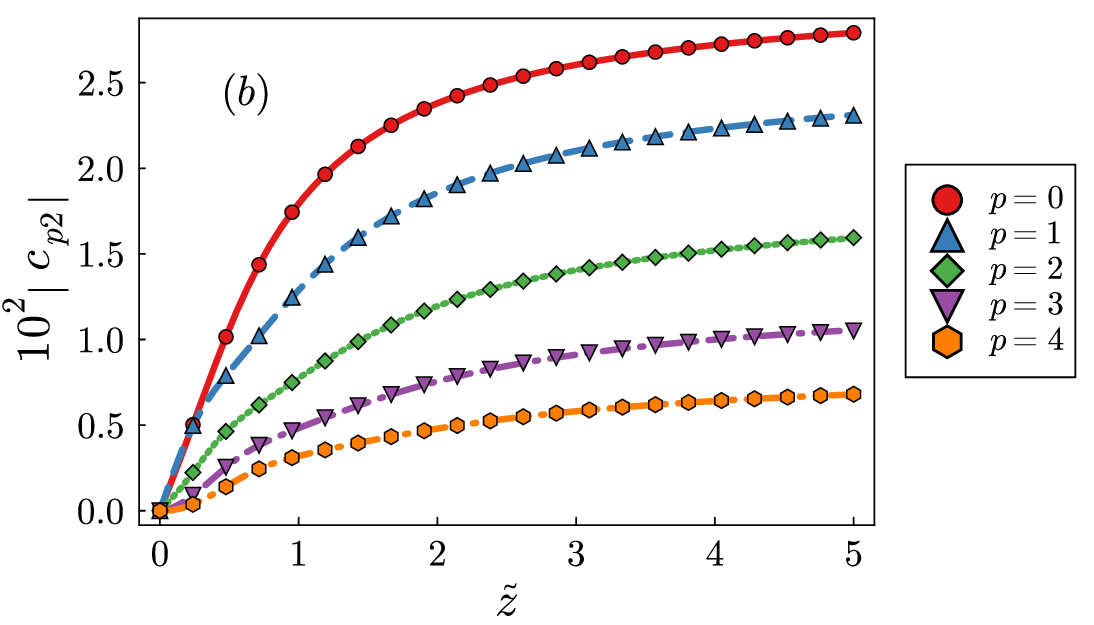}
        \caption{Overlap coefficients $c_{p2}$ for $g = 0.01$ and $l_0 = 2\,$. In plot $(a)$, we illustrate the short range propagation, where the solid lines represent the approximation given in Eq. \eqref{eq: cpl linear}. The appearance of radial modes with $0\leq p\leq 2$ is clear. In plot $(b)$ we show the diffraction zone, where more radial modes are generated. In both cases lines represent the approximation given by Eq. \eqref{eq: cpl nonlinear}
        and markers represent the results obtained from numerical integration of the nonlinear wave equation for $p=0$ (red solid line and circles), $p=1$ (blue dashed line and up triangles), $p=2$  (green dotted line and diamonds), $p=3$  (purple dash-dotted line and down triangles) and $p=4$  (orange dashed-double-dotted line and hexagons).}
        \label{fig:g=0.01}
    \end{figure}

\begin{figure}[ht]
%        \centering
        \includegraphics[width=\linewidth]{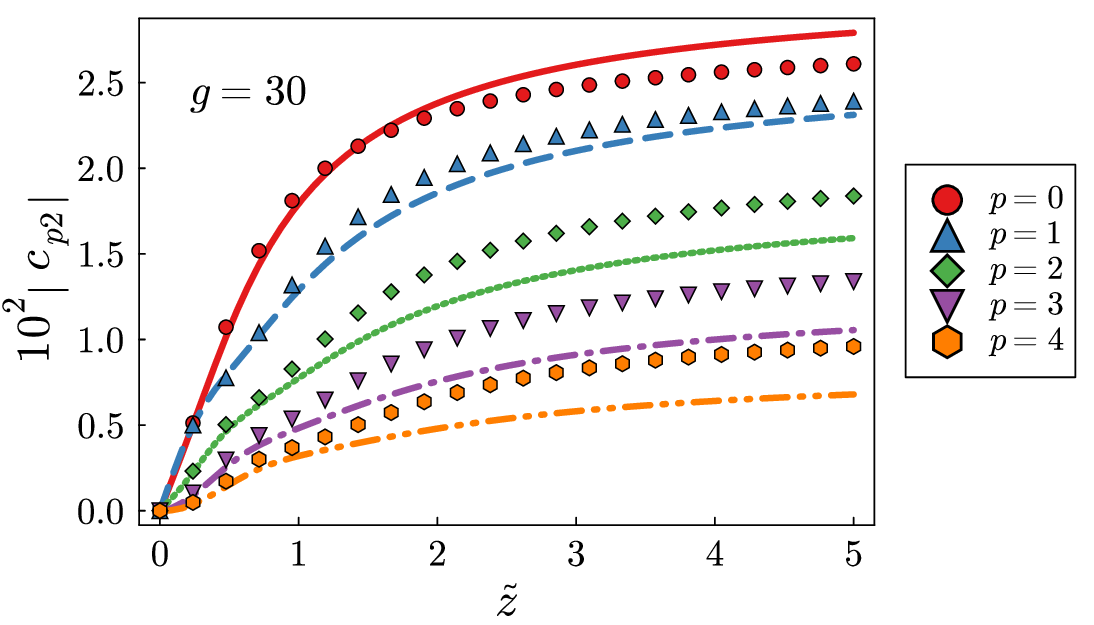}
        \includegraphics[width=\linewidth]{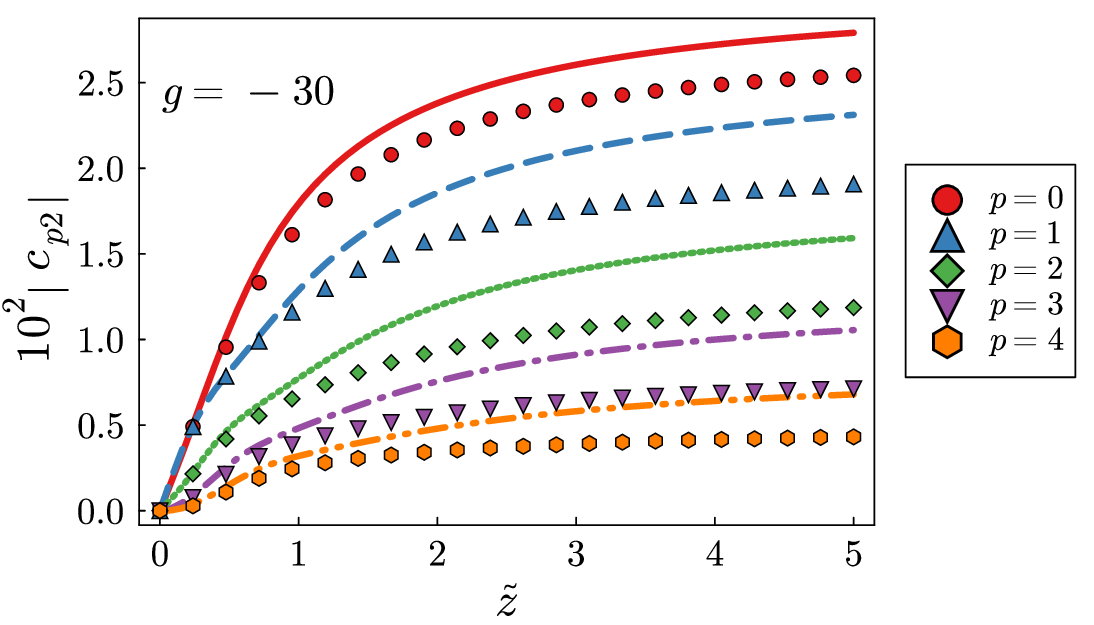}
        \caption{Overlap coefficients $c_{p2}$ for $g = \pm 30\,$. The solid lines represent the approximation given by Eq. \eqref{eq: cpl nonlinear}.
        In both cases lines represent the approximation given by Eq. \eqref{eq: cpl nonlinear}
        and markers represent the results obtained from numerical integration of the nonlinear wave equation for $p=0$ (red solid line and circles), $p=1$ (blue dashed line and up triangles), $p=2$  (green dotted line and diamonds), $p=3$  (purple dash-dotted line and down triangles) and $p=4$  (orange dashed-double-dotted line and hexagons).}
        \label{fig:g=+-30}
    \end{figure}

    Our numerical calculations will cover the two regimes corresponding to the focal ($z\ll z_R$) and diffraction 
    zones ($z\gtrsim z_R$). The values assumed for the nonlinear coupling $g$ will fall in the physical range given
    by Table \ref{tab:characteristic power}. In Fig. \ref{fig:g=0.01}, we plot the results for $g=0.01$ and $l_0=2$ 
    in the focal and diffraction zones. 
    This corresponds to a $1550nm$ wavelength laser with $650mW$ power propagating 
    in fused silica, for example. For the diffraction zone solution, the summation over $r$ was truncated at 
    $r_{max} = 10^3 \,$.
    One can easily see the development of a finite radial spectrum in the focal 
    zone, with $0\leq p\leq 2\,$, while other radial modes appear in the diffraction zone. 
    The results obtained from the perturbative 
    expressions are in excellent agreement with those given by numerical integration of the nonlinear wave equation. 
    Interestingly, the perturbative approach works well for $g\sim 1\,$, with significant deviations appearing 
    for $g\gtrsim 10\,$, as shown in Fig. \ref{fig:g=+-30}. Note that $\delta\psi$ defined 
    in Eq. \eqref{eq:deltapsi} is insensitive to the sign of $g$ up to first order perturbation, so that discrepancies 
    only appear for $g\gtrsim 10\,$, where second order corrections become important. These results may be 
    easily investigated with reasonable powers in fused silica, for example. 
    Moreover, in atomic vapors both the magnitude and sign of $g$ can be varied \cite{Kaiser2018}. 

\begin{figure}[ht]
%        \centering
        \includegraphics[width=\linewidth]{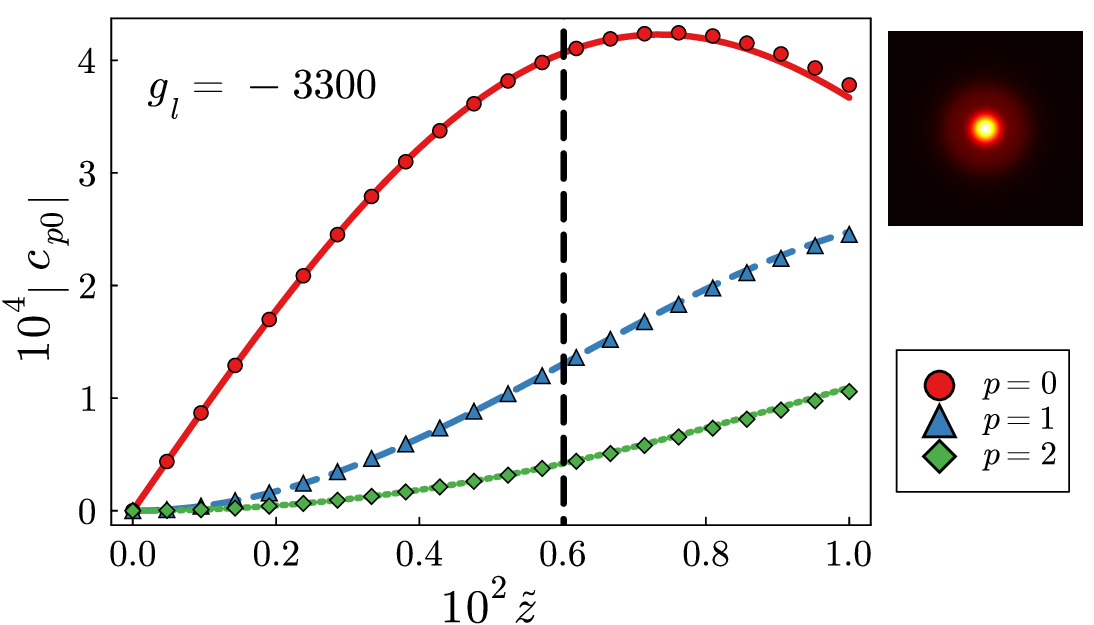}
        \includegraphics[width=\linewidth]{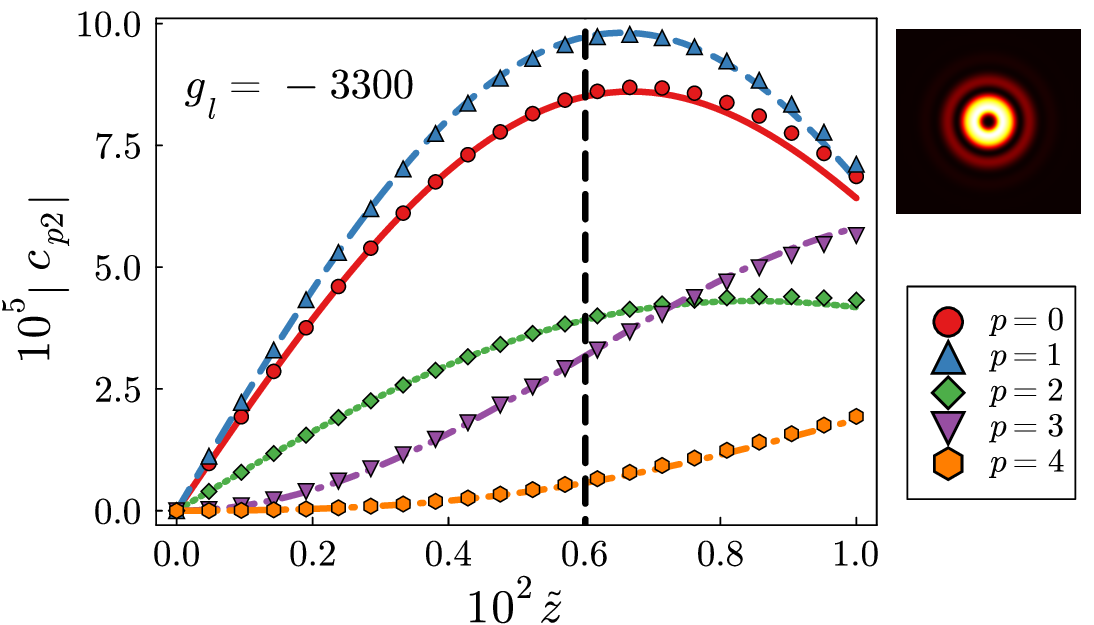}
        \caption{Overlap coefficients for $g_l = -3300\,$, corresponding to the experimental conditions of Ref. \cite{Kaiser2018}. Top: $l_0 = 0$ ($c_{p0}$).
        Bottom: $l_0 = 2$ ($c_{p2}$). The vertical dashed line indicates the interaction length in the experiment. The insets show the simulated far field images.
        In both cases lines represent the approximation given by  Eqs. \eqref{eq:spm-initial-solution} and \eqref{eq:spm-phase-solution}
        and markers represent the results obtained from numerical integration of the nonlinear wave equation for $p=0$ (red solid line and circles), $p=1$ (blue dashed line and up triangles), $p=2$  (green dotted line and diamonds), $p=3$  (purple dash-dotted line and down triangles) and $p=4$  (orange dashed-double-dotted line and hexagons).}
        \label{fig:kaiser}
    \end{figure}

    The appearance of rings with Gaussian beams in self phase modulation was thoroughly investigated in Ref. \cite{Cerda2010}. It was later 
    used to measure the nonlinear response of an atomic vapor \cite{Kaiser2018}. However, these studies did not consider the effects caused 
    by the orbital angular momentum. We next show numerical simulations of the radial mode generation under the physical conditions used in \cite{Kaiser2018}. 
    In Fig. \ref{fig:kaiser} we show the numerical results for the expansion coefficients of $\delta\psi$ for $l_0=0$ and $2$ under the experimental conditions. 
    The values of the coupling parameter $g$ were set to correspond to the same peak intensity for all OAM values. 
    It is straightforward to demonstrate that for a fixed beam power $P\,$, the peak intensity over the transverse plane is given by
    \begin{eqnarray}
    I^l_{max} = I_0\,
    \frac{\abs{l}^{\abs{l}}\,e^{-\abs{l}}}{\abs{l}!} 
    \approx \frac{I_0}{\sqrt{2\pi\abs{l}}}\;,
    \end{eqnarray}
    where the approximation follows from Stirling's formula and $I_0 = 2P/(\pi w_0^2)$ is the peak intensity of a Gaussian beam.
    Therefore, it will be useful to define 
    \begin{eqnarray}
    g_l = g \, \dfrac{\abs{l}^{\abs{l}}\,e^{-\abs{l}}}{\abs{l}!} 
    \approx \dfrac{g}{\sqrt{2\pi\abs{l}}}\,,
    \end{eqnarray}
    where the final approximation holds for $\abs{l}>0\,$.
    A meaningful comparison between the nonlinear effects on different OAM values must involve similar values of $g_l$ rather than $g\,$.
    For $l=0$ the radial mode components appear only in the non perturbative region and add up to the input mode, forming the far field ring structure.
    For $l=2$ the radial modes appear already in the perturbative region, causing a more pronounced ring structure in the far field.
    One can easily see that only the radial modes with $0\leq p\leq \abs{l_0}$ have nonzero derivative at $z=0\,$, as expected 
    from the perturbative solution in the focal zone.

    %

        %%%%%%%%%%%%%%%%%%%%%%%%%%%%%%%%%%%%%%%%%%%%%%%%%%%%%%%%%%%%%%%%%%%%%%%%%%%%%%%%%%%%%%%%%%%%%%%%%%%%%%%%%%%%%%%%%
	
	\section{Conclusion}
	\label{sec:conclusion}
	
	In this work we analyze the generation of a radial mode structure along propagation of an optical vortex 
	inside a nonlinear Kerr medium. Our analysis is based on a perturbative approach, useful for sufficiently 
	small nonlinear interaction. In this regime, the radial mode structure appears in the short range according 
	to a simple selection rule, once the waist of the mode basis is properly set. Radial numbers ranging between 
	zero and the absolute value of the input topological charge are generated. In the diffraction zone, the radial 
	mode intensity distribution is significantly affected by the Gouy phase match, an effect also observed in 
	intracavity parametric down conversion \cite{Alves2018} and four-wave mixing \cite{Sonja2021}. Our analytical 
	results obtained with the perturbative solution are legitimated by the numerical integration of the nonlinear wave 
	equation. Our results can be useful for a wide variety of physical systems displaying self-phase modulation, 
	such as optical propagation in atomic vapors and optical fibers. These methods can also be useful to describe 
        the evolution of vortex structures in Bose-Einstein condensates. A promising 
	lead to the present project is the investigation of quantum correlations associated with the radial 
	mode generation in the self-modulation process. The methods presented here will serve as the basis of 
	a quantum description, where the radial mode amplitudes are quantized.

	%%%%%%%%%%%%%%%%%%%%%%%%%%%%%%%%%%%%%%%%%%%%%%%%%%%%%%%%%%%%%%%%%%%%%%%%%%%%%%%%%%%%%%%%%%%%%%%%%%%%%%%%%%%%%%%%%
	
	\section*{Acknowledgments}
	
	Funding was provided by 
	Conselho Nacional de Desenvolvimento Científico e Tecnológico (CNPq), Coordenação de Aperfeiçoamento de Pessoal de Nível Superior (CAPES), 
        Fundação Carlos Chagas Filho de Amparo à Pesquisa do Estado do Rio de Janeiro (FAPERJ), Instituto Nacional de Ciência e Tecnologia de Informação Quântica (INCT-IQ 465469/2014-0) and Fundação de Amparo à Pesquisa do Estado de São Paulo (FAPESP, grant 2021/06823-5). A.Z.K. acknowledges financial support from Centre National de la Recherche Scientifique (CNRS) during his two-month stay at Laboratoire Kastler Brossel (LKB).
	
	%\bibliography{references.bib}

	%%%%%%%%%%%%%%%%%%%%%%%%%%%%%%%%%%%%%%%%%%%%%%%%%%%%%%%%%%%%%%%%%%%%%%%%%%%%%%%%%%%%%%%%%%%%%%%%%%%%%%%%%%%%%%%%%%%%%%
	\appendix
        \section{Conversion between mode families with different waists}
	\label{app:A}
	
	In this appendix we derive the transformation connecting two LG mode families with different waist parameters.
	It will help the identification of the optimal waist parameter for minimizing the number of radial modes in the short range.
	Let us consider two LG mode families $\{u_{pl}(\mathbf{r})\}$ and $\{\bar{u}_{pl}(\mathbf{r})\}\,$, with different 
	waist parameters and Rayleigh distances. Suppose their Rayleigh distances are related by $z_R = \eta \bar{z}_R\,$, which implies 
	$w_0^2 = \eta \bar{w}_0^2\,$. We want to determine the coefficients $\{\Lambda^{ml}_{qp}\}$ such that
	\begin{equation}\label{eq:uubar}
		\bar{u}_{qm}(\mathbf{r}) = \sum_{pl} \Lambda^{ml}_{qp}(\eta)\, u_{pl}(\mathbf{r})\,.
	\end{equation}
	The transformation coefficients are obtained by projecting Eq. \eqref{eq:uubar} onto the $\{u_{pl}(\mathbf{r})\}$ basis.
	Since the coefficients do not depend on the longitudinal coordinate $z\,$, we can calculate the transformation 
	coefficients at $z=0\,$. Therefore,
	%
%	\begin{widetext}
		\begin{eqnarray}\label{eq:G}
			&&\Lambda^{ml}_{qp}(\eta) = \int \bar{u}_{qm}(\boldsymbol{\rho},0)\, u^*_{pl}(\boldsymbol{\rho},0)\, d^2\boldsymbol{\rho}
			\nonumber\\
			&&= \frac{2}{\pi}\,\mathcal{N}_{qm}\mathcal{N}_{pl}\left(\sqrt{\eta}\right)^{\abs{m}+1}\!\!
			\int_{0}^{2\pi} e^{i(m-l)\phi}\, d\phi
			\\
			&&\times \int_{0}^{\infty} \!\!\!
			\left(2 \tilde{\rho}^2\right)^{\!\!\frac{\abs{m}+\abs{l}}{2}}\!\!
			L_{q}^{|m|}\!\left(2\eta\tilde{\rho}^2\right)\! L_{p}^{|l|}\!\left(2\tilde{\rho}^2\right)
			e^{-(\eta+1)\tilde{\rho}^2} \tilde{\rho} \,\,d\tilde{\rho}\,,
			\nonumber
		\end{eqnarray}
%	\end{widetext}
	%
	where $\tilde{\rho}=\rho^2/w^2_0\,$. Of course, the topological charges are not altered by a change in the 
	waist, so the angular integral imposes $l=m\,$, leading to
	\begin{equation}\label{eq:G2}
		\Lambda^{l}_{qp}(\eta) = \mathcal{N}_{ql}\mathcal{N}_{pl}\,\lambda_{qp}^{l}(\eta)\,,
	\end{equation}
	where we dropped one superfluous upper index in $\Lambda^{l}_{qp}$ and defined
	\begin{equation}
		\lambda_{qp}^{l}(\eta) = \left(\sqrt{\eta}\right)^{\abs{l}+1} \!\!\int_{0}^{\infty}\!\! x^{|l|}
		L_{q}^{|l|}\!\left(\eta x\right) \, L_{p}^{|l|}\!\left(x\right)
		e^{-\alpha x} dx \,,
	\end{equation}
	with $x=2\tilde{r}^2$ and $\alpha = (\eta + 1)/2\,$.
	The integral $\lambda_{qp}^{l}(\eta)$ can be obtained by using the generating function for the 
	generalized Laguerre polynomials,
	\begin{eqnarray}
		L_{p}^{|l|}\!\left(x\right) =
		\left.\frac{1}{p!}\frac{\partial^{p}}{\partial t^{p}}\left[\frac{e^{-xt/(1-t)}}{(1-t)^{|l|+1}}\right]\right|_{t=0}\,,
	\end{eqnarray}
	which leads to
	\begin{eqnarray}
		\lambda_{qp}^{l}(\eta) &=&
		\left.\frac{\left(\sqrt{\eta}\right)^{\abs{l}+1}}{q!p!} 
		\frac{\partial^{q}}{\partial t'^{q}}\frac{\partial^{p} F}{\partial t^{p}}
		\right|_{t,t'=0}\;,
		\\
		F(t,t') &=& 
		\frac{1}{\left[(1-t')(1-t)\right]^{|l|+1}}\int_{0}^{\infty}\!\! x^{|l|} e^{-b x} dx
		\nonumber\\
		&=& \frac{|l|!}{\left[b\,(1-t')(1-t)\right]^{|l|+1}}\;,
		\nonumber\\
		b &=& \frac{\eta\, t'}{1-t'} + \frac{t}{1-t} +\alpha\;.
		\nonumber
	\end{eqnarray}
	After a straightforward algebra, we arrive at
	\begin{eqnarray}
		\Lambda_{qp}^{l}(\eta) \!\!&=&\!\!(-1)^{p}\,\mathcal{N}_{ql}\mathcal{N}_{pl}
		\left(\frac{2\sqrt{\eta}}{1+\eta}\right)^{\abs{l}+1}
		\label{eq:lambdafinalz=0}\\
		&\times& \!\!\sum_{n=0}^{p} (-1)^n
		\frac{(q + p + |l| - n)!}{n! (p - n)! (q - n)!} \, 
		\left(\frac{1-\eta}{1+\eta}\right)^{q+p-2n} .
		\nonumber
	\end{eqnarray}
	In the main text we will use this mode conversion formula with $\eta=3$ to determine the overlap 
	integral.

\bibliography{bibfile}	

\end{document}